\documentclass[conference]{IEEEtran}
\IEEEoverridecommandlockouts
\usepackage{cite}
\usepackage{amsmath,amssymb,amsfonts}
\usepackage{algorithmic}
\usepackage{graphicx}
\usepackage{textcomp}
\usepackage{xcolor}
\def\BibTeX{{\rm B\kern-.05em{\sc i\kern-.025em b}\kern-.08em
    T\kern-.1667em\lower.7ex\hbox{E}\kern-.125emX}}

\makeatletter

\let\old@ps@IEEEtitlepagestyle\ps@IEEEtitlepagestyle

\def\confheader#1{%
    \def\ps@IEEEtitlepagestyle{%
        \old@ps@IEEEtitlepagestyle%
        \def\@oddhead{\strut\hfill#1\hfill\strut}%
        \def\@evenhead{\strut\hfill#1\hfill\strut}%
    }%
    \ps@headings%
}
    
\makeatother
    \confheader{%
    \parbox{20cm}{Accepted by Workshop on Digital Twins over NextG Wireless Networks, IEEE Global Communications Conference 2024 \\(IEEE GLOBECOM), \textcopyright 2024 IEEE}
}
\begin{document}

\title{Generative AI-enabled Digital Twins for 6G-enhanced Smart Cities}

\author{\IEEEauthorblockN{Kubra Duran\IEEEauthorrefmark{1}, Lal Verda Cakir\IEEEauthorrefmark{1},
Mehmet Ozdem\IEEEauthorrefmark{2}, Kerem Gursu\IEEEauthorrefmark{3}, Berk Canberk\IEEEauthorrefmark{1}}
 \
\IEEEauthorblockA{\IEEEauthorrefmark{1}School of Computing, Engineering and The Built Environment, Edinburgh Napier University, UK}
\IEEEauthorblockA{\IEEEauthorrefmark{2}Turk Telekom, Istanbul, Türkiye}
\IEEEauthorblockA{\IEEEauthorrefmark{3}BTS Group, Istanbul, Türkiye}
\\
		Emails: \{kubra.duran, lal.cakir, b.canberk\}@napier.ac.uk, mehmet.ozdem@turktelekom.com.tr, kerem.gursu@btsgrp.com}

\maketitle

\begin{abstract}
6G networks are envisioned to enable a wide range of applications, such as autonomous vehicles and smart cities. However, this rapid expansion of network topologies makes the management of 6G wireless networks more complex and leads to performance degradation. Even though state-of-the-art applications on network services are providing promising results, they also risk disrupting the network's performance. To overcome this, the services have to leverage what-if implementations covering a variety of scenarios. At this point, traditional simulations fall short of encompassing the dynamism and complexity of a physical network. To overcome these challenges, we propose the Generative AI-based Digital Twins. For this, we derive an optimization formula to differentiate different network scenarios by considering the specific key performance indicators (KPIs) for wireless networks. Then, we fed this formula to the generative AI with the historical twins and real-time twins to start generating the desired topologies. To evaluate the performance, we implement network and smart-city-oriented services, namely massive connectivity, tiny instant communication, right-time synchronization, and truck path routes. The simulation results reveal that our approach can achieve 38\% more stable network throughput in high device density scenarios.  Furthermore, the generated scenario accuracy is able to reach up to 98\% level, surpassing the baselines. 
\end{abstract}


\section{Introduction}

Digital Twins (DT) has become a key factor for enabling Next Generation (NextG) network services and meeting the desired network performance \cite{dt-6g}. The current developments in DT combined with Artificial Intelligence (AI) algorithms for wireless networks have shown significant performance improvements \cite{tunc_2024}. However, they also risk losing efficiency and disrupting the network's performance. This risk become highly prevalent, especially in 6G-enhanced smart cities \cite{aot}, because they rely on ultra connectivity of the Internet of Things (IoT) in real-time. Furthermore, massive machine-type communications (mMTC) are needed to achieve ultra connectivity characteristics in 6G wireless networks\cite{9686053}. In addition to this, the network must carry a large volume of data with low latency to make and apply decisions promptly. This requires the 6G-enhanced smart cities to have Tiny Instant Communication (TIC) service \cite{ecitygml}, \cite{comm-6g}.

At this point, several challenges must be addressed for digital twining in wireless networks in the context of 6G smart cities. In this study, we focus on two specific challenges as given below:

\begin{itemize}
    \item \textit{Degradation in network throughput with the enlarging wireless network topologies:} As wireless network topologies expand to cover larger areas and support increasing numbers of devices, network throughput experiences significant degradation \cite{access}. One of the main reasons for this is the capability of the network infrastructure to afford high device density levels. At this point, intelligent mechanisms should be developed to manage a large number of devices. 
    \item \textit{Lack of what-if implementation, and thus a low level of scenario accuracy}: The traditional simulations fall short of capturing the dynamism and complexity of the real world \cite{simdt}. In addition, changing set of parameters within a simulation scenario to see the future outcomes is not possible with the traditional methods. Therefore, accurate models that can reflect the potential outcomes of various scenarios on network performance are necessary.
\end{itemize}

Addressing the mentioned challenges is important for the full potential of smart city digital twins, thus achieving the vision of intelligent and responsive 6G-enhanced smart city environments. At this point, generative AI offers a key capability to create scenario twins. By learning the underlying structure and patterns of parameters, they can create scenarios that reflect potential real-world conditions. These scenario twins can simulate a wide range of situations for decision-making and service testing.

\section{State of the Art}

The digital twin technology have become a key element to develop 6G-enhanced smart cities \cite{9923927}. They have been used in different aspects, covering communication, security and intelligent services \cite{dt_enabling_challenges_open}. Real-time data combines AI and machine learning (ML) methods to apply adaptive management. For instance, in \cite{icc}, an adaptive synchronization was designed to increase the accuracy of the DT using reinforcement learning (RL). However, these adaptive control mechanisms also risk disrupting the ongoing operations in the physical networks \cite{wiley} when decisions are applied directly. To solve this, creating twins for the reserved usage of what-if capabilities is crucial. DTs have used data-driven models to create virtual replicas of the physical twins \cite{smart-city} via container-based deployments \cite{kubetwin}, and to emulate the distributed applications \cite{jaime}. Moreover, a DT-native modelling and communication framework is proposed in \cite{globecom}. However, these approaches have presented challenges in configuring the model-driven methods and acquiring the required data to ensure the data-driven methods converge \cite{wireless_ndt_genai}. 

At this point, generative AI methodologies can create a synthetic generation of scenarios by learning the underlying structure \cite{gai-survey}. With this approach, a conditional tabular generative adversarial network (CTGAN) has been used to generate DT data synthetically \cite{ak2024whatifanalysisframeworkdigital}. Moreover, different GAI algorithms can be deployed to generate data on the network's different services. However, this requires different models to effectively cooperate, which is an open issue \cite{genai-enabler}. Furthermore, while the benefits of this data augmentation approach expand the dataset, it cannot generate data reflecting the configuration changes. Considering these, in this article, we explore Large-Language Models (LLMs) to create scenarios and configure the simulation environment. Despite significant advancements in LLMs, their usage in DTs has been highly limited. In one of the works, LLMs are used to create an Asset Administration Shell (AAS) for semantic interoperability\cite{llm-aas}. Moreover, a dynamic simulation of the Heating, Ventilation and Air-Conditioning (HVAC) system has been implemented using LLMs in \cite{Yang_Siew_Joe-Wong_2024}. These works have showcased LLM's effectiveness in interpreting technical concepts, increasing efficiency and improving adaptability. 

As summarized above, several efforts have been made to address the ultra-massive connectivity and timely communication challenges in 6G smart city digital twins. However, none of them can handle all these challenges in a unified approach. Therefore, our research question in this study is,  \textit{“How can we design an effective Digital Twin framework for 6G wireless networks to perform intelligent scenario generation by enabling differentiated what-if testing and stable throughput for network and smart city services before deploying them in the real world?”} To hit this, we propose a generative AI-enabled digital twin framework with a novel optimization formula to differentiate between wireless network scenarios. We list the contributions of this study below:
\begin{itemize}
    \item We form a novel optimization formula to differentiate various wireless network scenarios. In this, we consider four significant KPIs of wireless networks: device density, packet deadlines, latency, and buffer size. 
    \item We design a Generative AI-enabled scenario twin creation framework fed by the historical twins, real-time twins, and derived optimization formula. Also, we consider three distinct wireless network scenarios as the generation criteria. 
    \item We test the proposed framework via network-oriented and smart-city-oriented services within the Twin Service Layer of the digital twin. As network-oriented services, we consider massive connectivity, tiny instant communication, and right-time synchronization services whereas evaluating truck routing as a smart-city service.
\end{itemize}

The remainder of the article is organized as follows: Section III explains the proposed generative AI-enabled digital twins framework by giving specific scenarios and services. Section IV gives the performance results of the proposed model. Finally, Section V concludes the paper.
\begin{figure*}[htbp]
\centerline{\includegraphics[width=0.99\linewidth]{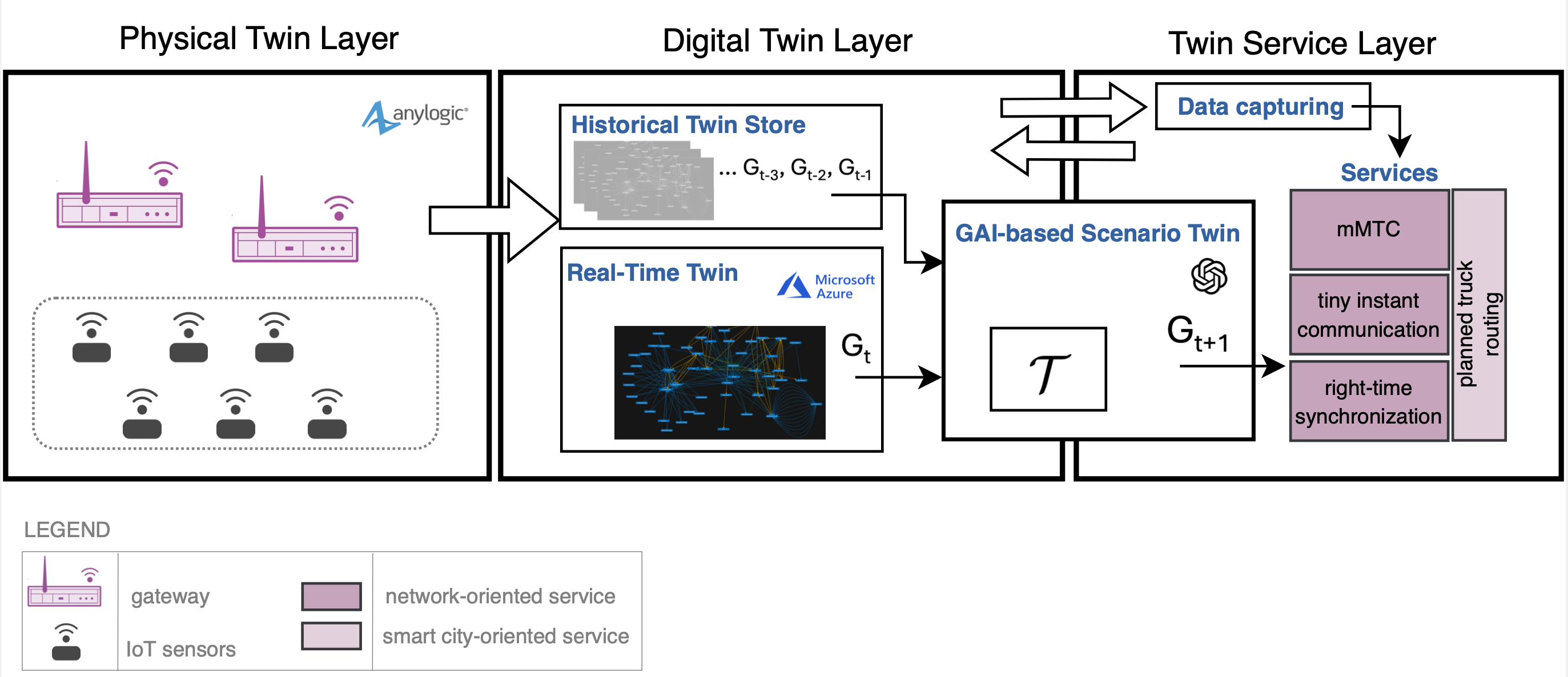}}
\caption{Generative AI-enabled Digital Twin Framework for 6G wireless network management.}
\label{fig}
\end{figure*}

\section{Generative AI-enabled Digital Twins Framework}
The proposed Generative AI-enabled Digital Twin framework consists of three distinct layers: The Physical Twin Layer, the Digital Twin Layer, and the Twin Service Layer. Each is specialized to form an end-to-end management approach for 6G-enhanced smart cities. The details of these layers are explained below.

\subsection{Physical Twin Layer}
Our main focus in this study is to test wireless networks' scalability, proactive responsiveness, and accuracy against changing scenarios. In our model, we have an increasing number of IoT sensors and gateways to be used in smart city services.  IoT sensors do not have mobility, and they are connected to one gateway during each scenario run. 

We conceptualize the Physical Twin Layer of the GAI-enabled Digital Twin framework by using an AnyLogic simulation environment. In this simulation, we create a practical replica of a physical IoT sensor network to test the defined network scenarios and services in the Twin Service Layer. 

\subsection{Digital Twin Layer:}
   This layer creates and manages virtual copies (mirrors) of physical objects and systems across the IoT wireless network. The quality of the twin modelling depends on the physical-to-virtual (P2V) communication link between physical and virtual counterparts. Our proposed Digital Twin Layer performs three types of twin creation, as explained below.
     
\subsubsection{Real-Time Twins}
We create the Real-Time Twins using Microsoft Azure Digital Twins (ADT) \cite{azure}. In this context, the network and the smart city are modelled in a graph formation, denoted as $G_{t}$ by utilizing the Digital Twin Definition Language (DTDL). Here, DTDL allows the precise modelling of physical and virtual elements within the IoT-based smart city network. It comprises four tuples, as explained below:

\begin{itemize}
    \item \textit{Property:} We use properties to store static and dynamic data of the Physical Twin Layer. Here, static data includes attributes like building dimensions or fixed infrastructure characteristics. In contrast, dynamic data corresponds to the measurements of IoT devices within the smart city services and the network measurements. 
    \item \textit{Telemetry:} They are used to initiate actions over the nodes in the ADT. This is done using the Telemetry API to send the related instructions to the twin within the telemetry data. This is then captured as an event at the ADT and routed to the endpoints asynchronously. This mechanism realizes the control of the nodes in the IoT network and the ADT. For instance, when the GAI-based scenario twin wants to retrieve a node's parameter, it uses telemetry.
    \item \textit{Component:} These individual elements within the twins are not managed independently but are integral to the overall system. This can be exemplified by the twin of the battery within the IoT device.
    \item \textit{Relationship:} Relationships define the physical and semantic connections between various smart city IoT network entities. These connections are represented by specific names in the DTDL. 
\end{itemize}

\subsubsection{Historical Twin Store} The historical data generated by IoT devices is stored using the Data History Connection in Azure Data Explorer (ADX). This storage ensures that the digital twins' historical state representations and associated time-series data are preserved until the current state. To indicate the data stored in this store, we use $\{G_{t-m}, ... G_{t-2}, G_{t-1}\}$. Here, $m$ depends on the implemented scenario and the capacity of the utilized datastore. ADX's time-series datasets are efficiently indexed and structured for optimized retrieval and query performance. Moreover, this is used to analyse and train AI/ML models in the service layer \cite{tgcn}.

\subsubsection{GAI-enabled Scenario Twin} 
In this module, we generate the possible future twin models of the IoT network by considering the differentiated scenarios and the desired key performance indicators (KPIs) respective for each scenario. Namely, scenario twins are created by taking the required scenario information as the input, and then the next-state digital twins are created as the output. This is implemented via the LLM (ChatGPT) console of our simulation environment (AnyLogic). The scenario twins allow us to perform various network scenarios to test the differentiated services within the Twin Service Layer. 

In this module, we define three distinct scenarios to be configured and generated, all of which are leveraging critical phases of the wireless network topologies. The details of these scenarios are explained in below. We define their KPI-based specific requirements to be fed into the generative AI. In this circumstance, the KPI set of scenarios, $K$, consists of four metrics, such as device density ($\rho$), packet deadlines ($d$), latency ($l$), and buffer size ($\alpha$). In each run, all KPI values are calculated \((\mathcal{T}_\rho, \mathcal{T}_d, \mathcal{T}_l, \mathcal{T}_\alpha)\), and relatively weighted according to the prioritization of the scenarios as given below:

\begin{equation}
\begin{cases}
    \ \ p(w_\rho, w_d) , \ \ D > D_{th}\\
    \ \  p(w_l, w_d) , \ \ L < L_{th}\\
    \ \  p(w_\rho, w_\alpha), \ \ A > A_{th}
\end{cases}
\label{eq1}
\end{equation}

In (\ref{eq1}), $p(.)$ function represents the $prioritization$, and it performs weight calculation. This function works based on assigning larger weight values to the arguments it takes. If there is no prioritization, then the weight values are randomly assigned for the scenario run. Moreover, $D$ implies the device density to be applied in the implemented service, $D_{th}$ the threshold value for the device density. Similarly, $L$ implies maximum tolerable latency in the scenario to be implemented, $L_{th}$ the threshold value for the latency. Moreover, $A$ implies the required accuracy in the service to be implemented, $A_{th}$ the threshold value for the accuracy value. The first case of this formula checks if the device density in the simulated scenario exceeds the threshold value. If it exceeds, the scenario is considered as a large topology and should be managed by prioritizing the (device density and packet deadline) parameters. Similar approaches run in the second and third cases by prioritizing (latency, packet deadline) and (device density, buffer size) parameters, respectively. We then form our objective function, $O$, to evaluate the wireless network scenarios and their corresponding KPI metrics in the same function, as given below:

\begin{equation}
O= w_{\rho}\mathcal{T}_\rho + w_{d}\mathcal{T}_d + w_{l}\mathcal{T}_l + w_{\alpha}\mathcal{T}_\alpha \\ \label{eq2}
\end{equation}

In (\ref{eq2}), each calculated KPI value and their corresponding weights calculated by the prioritization function are considered contributors to the main objective function, which will be considered the main network management function. In this circumstance, each prioritization forms an optimization task for each scenario to maximize the efficiency of the service while meeting the KPIs.

\begin{alignat}{4}
\label{eq3}
\text{Maximize}  & \quad &  O = \sum_{i} {w_{i}}{\mathcal{T}_i}   &&          & \quad \forall_{i} \in K\\
\label{eq4}
\text{subject to} & &\sum_{i} w_{i} && =1 & \\
\label{eq5}
&& p(w_x, w_y) &&  & \quad \forall_{(x,y)} \in K\\
\label{eq6}
  &&    w_x, w_y > 0 \\
 \label{eq7}
  && w_x, w_y > w_z, w_t  &&  & \quad  \forall_{(z,t)} \in K \wedge \neq (x, y)
\end{alignat}

We form our objective function in (\ref{eq3}). This function sums up the weighted KPIs of the network scenarios. The constraint (\ref{eq4}) ensures the sum of all weight values should be equal to one. (\ref{eq5}) denotes the prioritization function to be calculated depending on the implemented network scenario. Moreover, the weight values to be prioritized should be larger than zero (\ref{eq6}). Also, these prioritized weight values should not be equal to the other non-prioritized weight values (\ref{eq7}). In addition, the KPI set is denoted as \(K= \{\rho, d, l, \alpha\}\) in the mathematical notation. We give the details of each network scenarios below:

\begin{itemize}
    \item \textit{Scenario-1:} This is the \textit{base scenario} we consider that there is no specific demand for the network parameters. In this scenario, the network dynamics behave in a self-similar pattern. More specifically, no sudden changes result in peak points in the network topology and traffic. For this reason, prioritization is not performed for this scenario, and the weights for the network parameters are randomly assigned in this case. 
    \item \textit{Scenario-2:} This scenario represents \textit{high device density} cases, where an exponentially increasing number of IoT sensors are deployed within a specific area. This scenario simulates conditions that stress the scalability of the proposed digital twin framework on IoT networks. As the number of connected devices grows, the network must support this density without sacrificing performance \cite{bc1}, making scalability a critical consideration in wireless network management.
    \item \textit{Scenario-3:} This scenario is to address \textit{synchronization} challenge in digital twin-managed wireless networks. More specifically, capturing the data from a physical wireless network when triggered by a change in the topology preserves a significant amount of the network resources by presenting a resource-efficient digital twin framework. Therefore, the main target of this case is to measure \textit{“How accurate the twin models of the wireless sensors are?”} depending on the changing network circumstances.
\end{itemize}

\subsection{Twin Service Layer} 

The Twin Service Layer within the generative AI-enabled digital twin framework captures scenario-specific data and forms twin models. The details of this layer are given below:

\subsubsection{Data Capturing} Before starting the service implementations, this module collects service-based data from the Digital Twin Layer, especially from historical and real-time twin stores. Also, this module cleans data to prepare the data for service implementations.

\subsubsection{Services}
We selected four specific services to utilize and test our proposed generative AI-based digital twin framework, as detailed below:
\begin{itemize}
    \item \textit{mMTC:} This service supports the connectivity requirements of a vast number of IoT devices. For this service, we assume all IoT sensors deployed within the simulation area are identical but requesting UL messages with changing payload sizes. We also consider that IoT sensors indicate their UL requests simultaneously at 20\%, 45\%, and 70\% levels, respectively.  
    \item \textit{Tiny-instant communication (TIC):}  This service is based on our previous study, \cite{ecitygml}. In TIC service, the packet loss is desired to be nearly zero while responding to the IoT sensors' UL (uplink) and DL (downlink) message requests on time. The TIC service we considered in this study utilizes a Reinforcement Learning (RL)-based learner model to develop the most successive service.  
    \item \textit{Right-time synchronization:} As real-time synchronization requires high computational resources, right-time synchronization balances the benefits of real-time synchronization and resource usage (processing time, CPU usage) in wireless networks. In this service, we focus on the accuracy of the twin models by considering the right times while tracing the dynamic changes.
    \item \textit{Smart city application:} Besides the network infrastructure-focused services listed above, we also consider a smart city application scenario to investigate the success of GAI-based scenario twins and thus see its societal impact. For this reason, we consider the Planned Truck Routing (PTR) service based on our previous study \cite{t6conf} for waste management services in smart cities. The PTR service we consider within this study runs based on Support Vector Machines (SVM) to estimate the truck paths during the waste collection.  
\end{itemize}

\section{Performance Evaluation}
In this section, we investigate the performance of our proposed generative AI-enabled digital twin framework. For this, we measure (i) network throughput for the base and high device density scenarios according to enlarging topology sizes and (ii) accuracy of the generated scenario by using only historical twins, jointly using historical twins and real-time twins, and utilizing historical twins, real-time twins and generative AI together.

\textit{Experimental Setup:}
We use AnyLogic\textsuperscript{\copyright} to create our Physical Twin Layer of the proposed digital twin framework. We control generative AI-based scenario twin via the Anylogic Restfull API \cite{anylogic}. The simulation parameters used in this study are given in Table \ref{tab:sim}.
\def\arraystretch{1.5}
\begin{table}[thpb!]
    \centering
\caption{Simulation Parameters} %title of the table
\centering % centering table
\begin{tabular}{l c} % creating 2 columns
\hline % inserts single-line
Parameters&Values \\ [0.5ex]
\hline\hline %inserting double-line
Number of IoT sensors & \{50, 250, 1000\}\\
Number of gateways & \{2, 8, 20\}\\
Twinning rate & 0.8 \\
$D_{TH}, L_{TH}, A_{TH}$  & \{50, 0.9ms , 97\%\}  \\
Confidence interval  & 95\% \\
\hline % inserts single-line
\end{tabular}
\label{tab:sim}
\end{table}

\begin{figure}[htbp]
\centerline{\includegraphics[width=0.99\linewidth]{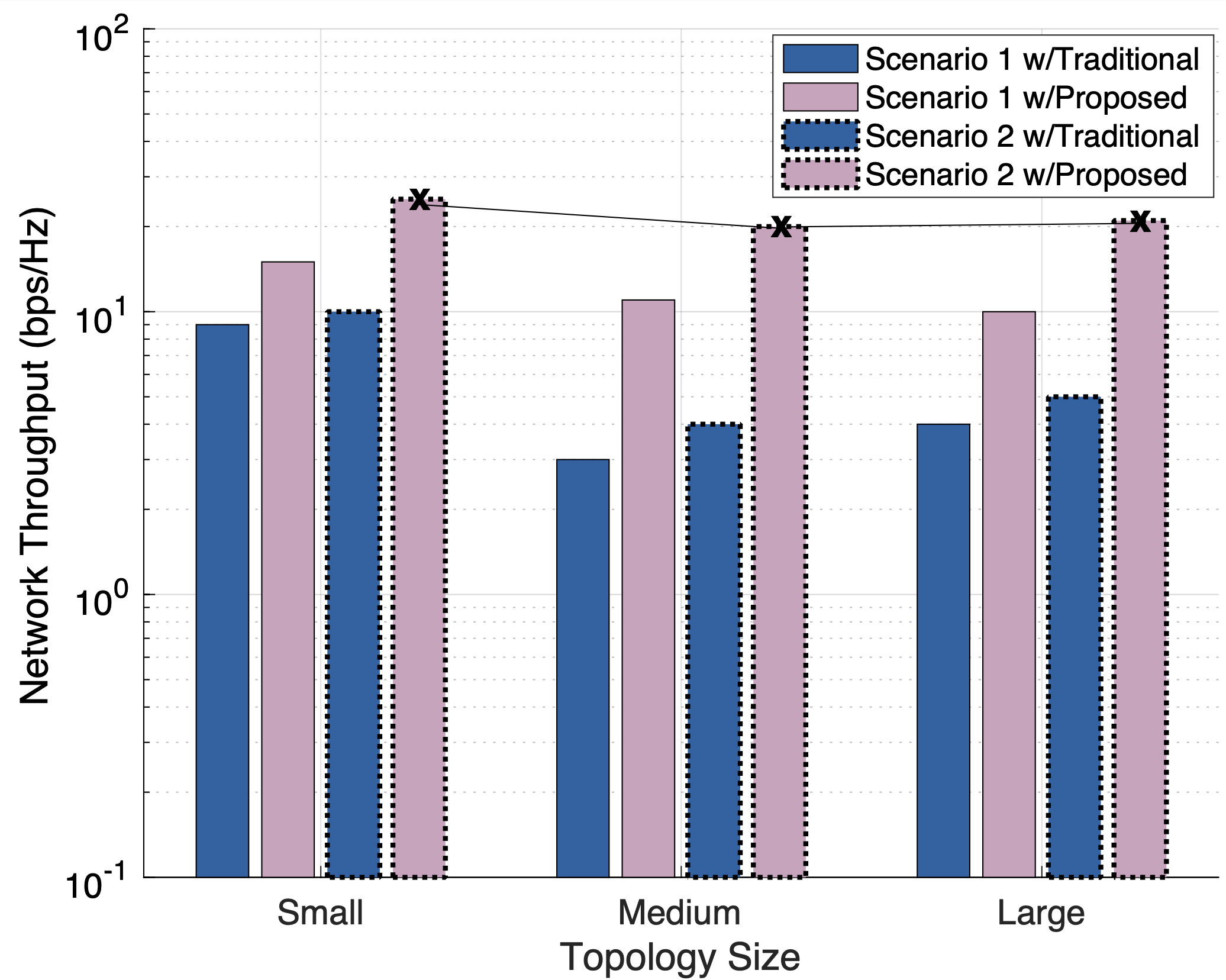}}
\caption{Network throughput comparison for Scenario-1 and Scenario-2 according to the changing topology sizes for generative AI-enabled digital twin (proposed) and traditional method.}
\label{figth}
\end{figure}

We start with testing the performance of our optimization formula. We define a high-density scenario and compare it with the baseline scenario (explained in Section III.B.3). In this simulation run, we test the performance of mMTC and TIC services. For this, we generated small, medium and large-scale IoT topologies using identical sensors and gateways. As given in \ref{tab2}, we create 50, 250, and 1000 sensors in the AnyLogic environment to represent different topology sizes. We also create 2, 8, and 20 gateways for the respective topologies. Also, we consider that UL packet sizes for IoT sensors are  2Kbps at maximum.  In Fig. \ref{figth}, “Scenario-1” presents the base scenario with the self-similar network pattern, while “Scenario-2” represents the high device density scenario. During our simulations, we run two individual scenarios for three different topology sizes with the traditional simulation method and the proposed generative AI-based digital twin framework. As a result of our simulations, we observe that the proposed generative AI-based digital twin framework achieves 38\% more stable network throughput on average. This is because our optimization formula prioritizes the KPIs of high device density scenarios and tries to cover all IoT sensors presented in the topology while maximizing network efficiency. On the contrary, as the base scenario does not require KPI prioritization, the weight values are assigned randomly in the run of this scenario using the traditional method. 

\def\arraystretch{1.5}
\begin{table}[htbp]
\caption{Topology Sizes}
\begin{center}
\begin{tabular}{|c|c|c|c|}
\hline
\textbf{}&\multicolumn{3}{|c|}{\textbf{Tested Topologies}} \\
\cline{2-4} 
\textbf{} & \textbf{\textit{small}}& \textbf{\textit{medium}}& \textbf{\textit{large}} \\
\hline
Device density$^{\mathrm{*}}$ & 50& 250& 1000 \\
\hline
Gateways & 2& 8& 20 \\
\hline
\multicolumn{4}{l}{$^{\mathrm{*}}$(number of devices/$m^2$).}
\end{tabular}
\label{tab2}
\end{center}
\end{table}

\begin{figure}[htbp]
\centerline{\includegraphics[width=0.99\linewidth]{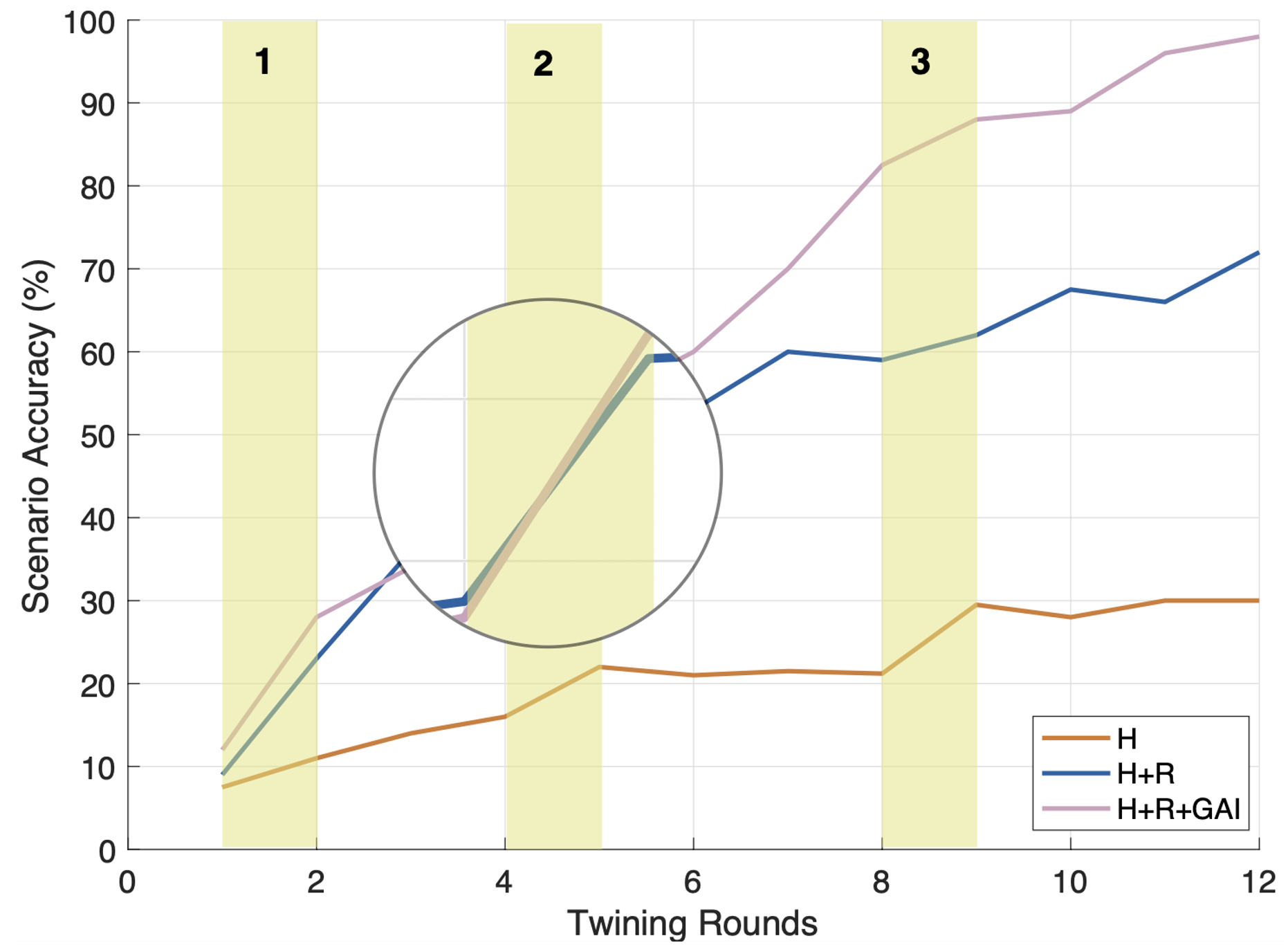}}
\caption{Scenario accuracy comparison for right-time synchronization service by using historical twins, real-time twins and generative AI.}
\label{fig3}
\end{figure}

In the second set of our experiments, we test the accuracy of the generated scenario for the right time synchronization and PTR smart city services. For this, we define our scenario as synchronization-oriented, meaning that the scenario requires high accuracy during the simulation run. We perform twelve consecutive twinning rounds with a constant twinning rate of 0.8. Here, we decide the total number of performed twining rounds by observing the scenario accuracy value. If we desire to see 100\% accuracy, then we can perform one more twinning round. As seen in Fig. \ref{fig3}, we generate the synchronization-oriented scenario as detailed in “Scenario-3” in Section III.B.3 by using only historical twins (shown as H with orange line), the hybrid usage of historical twins and real-time twins (shown as H+R with blue line), and also by utilizing our proposed method (shown as H+R+GAI with purple line). According to the simulation results, we highlight three regions with yellow in the figure to indicate the change in the behaviour of scenario accuracy. Namely, in the first region, all models show a linear increase, especially the results of the H+R and H+R+GAI methods, which are close to each other. In the second region, these two models catch each other, resulting in the same accuracy levels of around 50\%. This round (fifth) is also noted as nearly half of the twining progress. After this round, the proposed method starts to surpass the other methods. As seen in the third region, the proposed method reaches 90\% accuracy levels while others achieve 60\% most. Finally, the proposed twin generation method reaches 98\% accuracy levels at the end of the twining rounds. The main reason for this significant achievement is that our proposed method takes advantage of generative AI while utilizing a scenario-specific optimization function. In other words, the proposed generative AI-enabled digital twin framework takes criteria given as the inputs to the system, performs optimization and acts according to the desired system requirements. 

\section{Conclusion}
In this research, we propose the generative AI-enabled digital twins to address two specific challenges in 6G wireless networks: (i) throughput degradation and (ii) lack of what-if scenario implementation to enable highly accurate test scenarios. For this, we derive an optimization formula to differentiate wireless network scenarios. We then feed this formula along with historical and real-time twins to generate desired network topologies. Our proposed framework behaves as a finely tunable platform to generate differentiated scenarios for 6G wireless networks. Our simulation results demonstrate 38\% improvement in the network throughput under high device density conditions, and the generated scenario accuracy reaches 98\% levels.

As a future work, we plan to enhance our generative AI-based digital twinning framework by enlarging our KPI set to involve more wireless network services.

\section*{Acknowledgment}
This work was partially supported by The Scientific and Technological Research Council of Turkey (TUBITAK) 1515 Frontier R\&D Laboratories Support Program for BTS Advanced AI Hub: BTS Autonomous Networks and Data Innovation Lab. Project 5239903. The work of K. Duran and L.V. Cakir was partially supported by DeepMind. 

\bibliographystyle{IEEEtran}
\bibliography{IEEEabrv, references}

\begin{thebibliography}{10}
\providecommand{\url}[1]{#1}
\csname url@samestyle\endcsname
\providecommand{\newblock}{\relax}
\providecommand{\bibinfo}[2]{#2}
\providecommand{\BIBentrySTDinterwordspacing}{\spaceskip=0pt\relax}
\providecommand{\BIBentryALTinterwordstretchfactor}{4}
\providecommand{\BIBentryALTinterwordspacing}{\spaceskip=\fontdimen2\font plus
\BIBentryALTinterwordstretchfactor\fontdimen3\font minus \fontdimen4\font\relax}
\providecommand{\BIBforeignlanguage}[2]{{%
\expandafter\ifx\csname l@#1\endcsname\relax
\typeout{** WARNING: IEEEtran.bst: No hyphenation pattern has been}%
\typeout{** loaded for the language `#1'. Using the pattern for}%
\typeout{** the default language instead.}%
\else
\language=\csname l@#1\endcsname
\fi
#2}}
\providecommand{\BIBdecl}{\relax}
\BIBdecl

\bibitem{dt-6g}
N.~P. Kuruvatti, M.~A. Habibi, S.~Partani, B.~Han, A.~Fellan, and H.~D. Schotten, ``Empowering 6g communication systems with digital twin technology: A comprehensive survey,'' \emph{IEEE Access}, vol.~10, pp. 112\,158--112\,186, 2022.

\bibitem{tunc_2024}
C.~Tunc, T.~X. Tran, and K.~Joshi, ``{Digital Twins for Beyond 5G},'' in \emph{{AI in Wireless for Beyond 5G Networks}}, 1st~ed.\hskip 1em plus 0.5em minus 0.4em\relax CRC Press, 2024, pp. 169--190.

\bibitem{aot}
K.~Duran, M.~Özdem, T.~Hoang, T.~Q. Duong, and B.~Canberk, ``Age of twin (aot): A new digital twin qualifier for 6g ecosystem,'' \emph{IEEE Internet of Things Magazine}, vol.~6, no.~4, pp. 138--143, 2023.

\bibitem{9686053}
Y.~Li, X.~Jian, K.~Yu, N.~Kumar, and S.~Cai, ``Theoretical performance analysis of distributed queue for massive machine type communications: Throughput, latency, energy consumption,'' \emph{IEEE Transactions on Network and Service Management}, vol.~19, no.~2, pp. 818--828, 2022.

\bibitem{ecitygml}
K.~Duran, E.~Ak, G.~Yurdakul, and B.~Canberk, ``6g-enabled dtaas (digital twin as a service) for decarbonized cities,'' in \emph{2023 IEEE International Conference on Communications Workshops (ICC Workshops)}, 2023, pp. 421--426.

\bibitem{comm-6g}
H.~Viswanathan and P.~E. Mogensen, ``Communications in the 6g era,'' \emph{IEEE Access}, vol.~8, pp. 57\,063--57\,074, 2020.

\bibitem{access}
R.~Venanzi, L.~Foschini, P.~Bellavista, B.~Kantarci, and C.~Stefanelli, ``Fog-driven context-aware architecture for node discovery and energy saving strategy for internet of things environments,'' \emph{IEEE Access}, vol.~7, pp. 134\,173--134\,186, 2019.

\bibitem{simdt}
M.~Schluse and J.~Rossmann, ``From simulation to experimentable digital twins: Simulation-based development and operation of complex technical systems,'' in \emph{2016 IEEE International Symposium on Systems Engineering (ISSE)}, 2016, pp. 1--6.

\bibitem{9923927}
N.~P. Kuruvatti, M.~A. Habibi, S.~Partani, B.~Han, A.~Fellan, and H.~D. Schotten, ``Empowering 6g communication systems with digital twin technology: A comprehensive survey,'' \emph{IEEE Access}, vol.~10, pp. 112\,158--112\,186, 2022.

\bibitem{dt_enabling_challenges_open}
A.~Fuller, Z.~Fan, C.~Day, and C.~Barlow, ``Digital twin: Enabling technologies, challenges and open research,'' \emph{IEEE Access}, vol.~8, pp. 108\,952--108\,971, 2020.

\bibitem{icc}
L.~V. Cakir, K.~Duran, C.~Thomson, M.~Broadbent, and B.~Canberk, ``Ai in energy digital twining: A reinforcement learning-based adaptive digital twin model for green cities,'' \emph{arXiv preprint arXiv:2401.16449}, 2024.

\bibitem{wiley}
K.~Duran, B.~Karanlik, and B.~Canberk, ``Graph theoretical approach for automated ip lifecycle management in telco networks,'' \emph{Wiley Int J Network Mgmt}, vol.~31, no. 4, e2138, 2021.

\bibitem{smart-city}
N.~Mohammadi and J.~E. Taylor, ``Smart city digital twins,'' in \emph{2017 IEEE Symposium Series on Computational Intelligence (SSCI)}, 2017, pp. 1--5.

\bibitem{kubetwin}
D.~Borsatti, W.~Cerroni, L.~Foschini, G.~Ya~Grabarnik, L.~Manca, F.~Poltronieri, D.~Scotece, L.~Shwartz, C.~Stefanelli, M.~Tortonesi, and M.~Zaccarini, ``Kubetwin: A digital twin framework for kubernetes deployments at scale,'' \emph{IEEE Transactions on Network and Service Management}, vol.~21, no.~4, pp. 3889--3903, 2024.

\bibitem{jaime}
S.~Laso, L.~Martín, J.~L. Herrera, J.~Galán-Jiménez, J.~Berrocal, and J.~M. Murillo, ``Dantalion: Digital twinning the computing continuum,'' in \emph{2023 IEEE Globecom Workshops (GC Wkshps)}, 2023, pp. 1303--1306.

\bibitem{globecom}
K.~Duran, M.~Broadbent, G.~Yurdakul, and B.~Canberk, ``Digital twin-native ai-driven service architecture for industrial networks,'' in \emph{2023 IEEE Globecom Workshops (GC Wkshps)}, 2023, pp. 1297--1302.

\bibitem{wireless_ndt_genai}
Z.~Tao, W.~Xu, Y.~Huang, X.~Wang, and X.~You, ``Wireless network digital twin for 6g: Generative ai as a key enabler,'' \emph{IEEE Wireless Communications}, vol.~31, no.~4, pp. 24--31, 2024.

\bibitem{gai-survey}
A.~Celik and A.~M. Eltawil, ``At the dawn of generative ai era: A tutorial-cum-survey on new frontiers in 6g wireless intelligence,'' \emph{IEEE Open Journal of the Communications Society}, vol.~5, pp. 2433--2489, 2024.

\bibitem{ak2024whatifanalysisframeworkdigital}
\BIBentryALTinterwordspacing
E.~Ak, B.~Canberk, V.~Sharma, O.~A. Dobre, and T.~Q. Duong, ``What-if analysis framework for digital twins in 6g wireless network management,'' 2024. [Online]. Available: \url{https://arxiv.org/abs/2404.11394}
\BIBentrySTDinterwordspacing

\bibitem{genai-enabler}
Z.~Tao, W.~Xu, Y.~Huang, X.~Wang, and X.~You, ``Wireless network digital twin for 6g: Generative ai as a key enabler,'' \emph{IEEE Wireless Communications}, vol.~31, no.~4, pp. 24--31, 2024.

\bibitem{llm-aas}
Y.~Xia, Z.~Xiao, N.~Jazdi, and M.~Weyrich, ``Generation of asset administration shell with large language model agents: Toward semantic interoperability in digital twins in the context of industry 4.0,'' \emph{IEEE Access}, vol.~12, pp. 84\,863--84\,877, 2024.

\bibitem{Yang_Siew_Joe-Wong_2024}
\BIBentryALTinterwordspacing
H.~Yang, M.~Siew, and C.~Joe-Wong, ``An llm-based digital twin for optimizing human-in-the loop systems,'' 2024. [Online]. Available: \url{http://arxiv.org/abs/2403.16809}
\BIBentrySTDinterwordspacing

\bibitem{azure}
\BIBentryALTinterwordspacing
``\BIBforeignlanguage{en}{Azure digital twins}.'' [Online]. Available: \url{https://learn.microsoft.com/en-gb/azure/digital-twins/}
\BIBentrySTDinterwordspacing

\bibitem{tgcn}
K.~Duran and B.~Canberk, ``{Digital Twin Enriched Green Topology Discovery for Next Generation Core Networks},'' \emph{IEEE Transactions on Green Communications and Networking}, vol.~7, no.~4, pp. 1946--1956, 2023.

\bibitem{bc1}
D.~M. Gutierrez-Estevez, B.~Canberk, and I.~F. Akyildiz, ``Spatio-temporal estimation for interference management in femtocell networks,'' in \emph{2012 IEEE 23rd International Symposium on Personal, Indoor and Mobile Radio Communications - (PIMRC)}, 2012, pp. 1137--1142.

\bibitem{t6conf}
E.~Ak, K.~Duran, O.~A. Dobre, T.~Q. Duong, and B.~Canberk, ``T6conf: Digital twin networking framework for ipv6-enabled net-zero smart cities,'' \emph{IEEE Communications Magazine}, vol.~61, no.~3, pp. 36--42, 2023.

\bibitem{anylogic}
\BIBentryALTinterwordspacing
``Anylogic restfull api.'' [Online]. Available: \url{https://anylogic.help/cloud/api/rest.html}
\BIBentrySTDinterwordspacing

\end{thebibliography}

\end{document}